\documentstyle[12pt]{article}
\begin{document}

\author {Victor Novozhilov and Yuri Novozhilov  \\
V.A.Fock Department of Theoretical Physics, \\ St.Petersburg State University 198904,
St.Petersburg, Russia }
\title { Klein-Fock equation, proper-time formalism and symmetry of Hydrogen
atom
\footnote{To be published in the Proceedings of the UNESCO International School
of Physics "Quantum Theory '  part 2, commemorating the 100th anniversary
of the birth of V.A.Fock. St.Peterburg , SPbU, 1999. \\
This work was supported in part by RFBR (Grant 97-01-01186) and
by GRACENAS (Grant 97-0-14.1-61). \\
Email: yunovo@niif.spb.su, novozhil@nvv.gc.spb.ru}}
\date {}
\maketitle

{\bf Abstract}

We present main points of  some of Fock papers in Quantum Theory,
which were not properly followed when published.

\vskip 6pt
{\bf Introduction}
\vskip 3pt

Fock papers on Quantum Theory have quite different lifes in Physics. Some of
them were immediately recognized with Fock name attached to them.
Hartree-Fock approach, Fock space and Fock representation are very well
known examples. Another ones were not properly used  when published. To this  class
belong papers where Fock was ahead of time, and his comptemporaries were
not able to really understand their importance.
Klein-Fock equation, proper-time formalism and symmetry of Hydrogen atom are
best examples of Fock papers within such a class.

In this lecture we explain main points of these Fock papers and
remind Fock interpretation of energy-time uncertainty relation, as  well as his  ideas
on generalization of the concept of physical space published in his last
paper.

\vskip 6pt
{\bf Generalization of Schroedinger equation and Klein-Fock equation}
\vskip 3pt

V.A.Fock sent his first paper on Quantum Mechanics for publication in Zs.f.
Physik one week after Schroedinger's first paper had reached Leningrad. Fock
generalized Schroedinger equation for the case of magnetic field and
obtained an expression for the (normal) level splitting in a magnetic field
as well as level splitting of Hydrogen atom in an electric field [1].

In his second paper on Quantum Mechanics [2] V.A.Fock presented relativistic
generalization of Schroedinger equation for a particle in electromagnetic
field on curved space. O.Klein [3] has also derived a relativistic
generalization of Schroedinger equation and published it in Zs.f.Physik,
vol.37 before Fock. But the paper by Fock in vol.39 was received by this
journal before publication of Klein's paper. The third paper on the same
subject for the simplest case of a free partcle on a flat space was written
by W.Gordon [4]; it was received by Zs.F.Physik after paper by Klein was
published and appeared in vol.40. Nevertheless, in many textbooks the
equation for spin 0 particle is called as the Klein-Gordon equation.

One might see the reason for such discrimination of Fock in a character of
the paper [2]. This paper contains much more material than a simple
relativistic generalization of Schroedinger Eq. to Eq. for a free spin 0
particle. At that time, this material could not be appreciated by most
physicists.

Let us consider briefly main points of Fock's paper [2]. Fock introduces
five dimensional space with the metric depending on electromagnetic
potential $A_\mu $

\begin{eqnarray*}
d\sigma ^2 &=&g_{\mu \nu }dx_\mu dx_\nu -\frac{e^2}{m^2c^4}\left( A_\nu
dx_\nu +du\right) ^2 \\
&&
\end{eqnarray*}
where $u$ is an additional coordinate. In classical physics null geodesic
line $d\sigma =0$ describes a trajectory of charged particle in this space.
Corresponding action $S$ will have five dimensional gradient squared equal
to zero. Four dimensional action $W$ is related to $S$

\begin{eqnarray*}
S &=&\frac ecu+W \\
&&
\end{eqnarray*}
Five dimensional equations are invariant under transformation

\begin{eqnarray*}
A_\nu &=&A_\nu ^{\prime }+\partial _\nu f \\
u &=&u^{\prime }-f \\
&&
\end{eqnarray*}
which later [5] were called as {\bf gradient transformation}.

Both classical and quantum eqations act in the same space.Therefore, the
corresponding quantum wave equation for the wave function $\Psi $ is the
d'Alambert equation in five dimensional space

$$
\frac 1{\sqrt{-g}}\frac \partial {\partial x_\mu }\left( \sqrt{-g}g^{\mu \nu
}\frac{\partial \Psi }{\partial x_\nu }\right) -2A^\nu \frac{\partial ^2\Psi
}{\partial u\partial x_\nu }+\left( A_\nu A^\nu -\frac{m^2c^4}{e^2}\right)
\frac{\partial ^2\Psi }{\partial u^2}=0
$$
Four dimensional wave function $\psi $ can be found from $\Psi $ by a phase
transformation

$$
\Psi =e^{i\frac e{hc}u}\psi
$$
and five dimensional eq. for $\Psi $ produces immediately the Klein-Fock
equation for $\psi $ in electromagnetic field and on the curved space.

Gradient transformation of $A_\nu $ and $u$ induces transformation of the
wave function
\begin{eqnarray*}
\psi ^{\prime } &=&e^{-i\frac{ef}{hc}}\psi \\
&&
\end{eqnarray*}
This transformation rule of wave function under gauge tranformation was
first written in Fock's paper on the Klein-Fock equation.

These two papers of young Fock made his name known to theoreticians. He got
Rockefeller fellowship and went to study in Goettingen and Paris.

\vskip 6pt
{\bf Proper time formalism and Fock gauge condition (1937)}
\vskip 3pt

In this paper V.A.Fock [6] considers Dirac equation with an external field
and develops a relativistically and gauge invariant quasiclassical method of
integration. At that time, such a method was needed to deal with vacuum
polarization in the Dirac positron theory. It was shown by V.Weisskopf [7]
that logarithmic divergences appear in calculation of vacuum polarization.
The problem was how to separate finite parts from divergent expressions in a
unique relativistically and gauge invariant way.

Fock proposed to ensure invariance and uniqueness of calculations by using
invariant quantities only. He introduces an invariant parameter to consider
evolution in a symmetric manner with respect to all four relativistic
coordinates and shows how to relate new five dimensional space to the four
dimensional one in a unique way so that an additional parameter acquires a
meaning of proper time. Fock notes that working in five dimensional space is
necessary, because by this trick one gets uniqueness of divergent
expressions due to the fact that the Riemann fundamental solution can be
defined uniquely only in spaces of odd dimensions. Fock introduced also a
new gauge condition for electromagnetic potential without singularities
which enables to express potentials in terms of field strengths uniquely.

Fock considers second order Dirac equation in the four dimensional space

$$
\{P_\mu P^\mu +m^2c^2+\frac{eh}c\left( \overline{\sigma }\cdot {\bf H}
\right) -\frac{ieh}c\left( \overline{\alpha }\cdot {\bf E}\right) \}\Psi
=0
$$
or symbolically

$$
h^2\Lambda \Psi =0
$$
where $\Lambda $ is a second order operator. One can present a solution as
an integral

$$
\Psi =\int_CFd\tau
$$
over a function $F$ of five variables $x_\mu ,\tau $ with a given
integration path $C$ in variable $\tau $ . Then $F$ should be a solution of
Dirac equation with proper time $\tau $

$$
\frac{h^2}{2m}\Lambda F=ih\frac{\partial F}{\partial \tau }
$$
subject to condition

$$
\int_C\frac{\partial F}{\partial \tau }d\tau =F_C=0
$$
where $F_C$ is function $F$ taken at the path $C$ . Fock shows how to choose
the path $C$ to give $\tau $ a meaning of proper time.

All calculations should be done in five dimensional space where separation
of divergences is unique. Integration over proper time completes
calculations. Proper time integrals does not depend on gauge and are
relativistically invariant by their definition.

Fock proper time formalism was developed by J.Schwinger [8] in 1951 and by
B.S.De Witt [9] in 1965. It is usually referred to as Schwinger-De Witt
proper time method, although it was formulated by Fock. Fock paper was too
much ahead of time.

Fock gauge condition for electromagnetic potential reads

$$
(x_\nu -x_\nu ^0)A^\nu =0
$$
assuming that potential $A^\nu $ is non-singular. The potential can be
expressed in terms of field strengths $A_{\mu \nu }$ by averaging between
points $x$ and $x^0$ according to

$$
\stackrel{}{\overline{f}=2\int_0^1f\left[ x^0+\left( x-x^0\right) u\right]
udu} 
$$
Then

\begin{eqnarray*}
A^\nu &=&\frac 12\left( x-x^0\right) _\mu \overline{A}^{\mu \nu } \\
&&
\end{eqnarray*}
Fock gauge condition is especially useful when dealing with selfdual fields.
Fock condition has another advantage: as it was found later [10], in quantum
field theory the Faddeev-Popov ghosts decouple .

\vskip 6pt
{\bf Symmetry of Hydrogen atom and dynamical groups}
\vskip 3pt

According to Schroedinger equation, energy levels of a charge in a
spherically symmetric field are characterized by two quantum numbers:
principal quantum number and eigenvalue of angular momentum. However, in
Hydrogen atom energy levels depend on principal quantum number only. The
origin of degeneracy was guessed to be in additional symmetry of Hydrogen
atom. This problem was known long before Fock, but only Fock was able to
solve the problem. He did it in paper ''Hydrogen atom and non-Euclidean
geometry'' [11] in a simple and elegant way.

Fock starts by writing down the Schroedinger equation with Coulomb potential
in momentum space as an integral equation

$$
\frac 1{2m}p^2\psi \left( {\bf p}\right) =-\frac{Ze^2}{2\pi ^2h}\int 
\frac{\psi \left( {\bf p}^{\prime }\right) }{\left| {\bf p-p}^{\prime
}\right| }d {\bf p}^{\prime } 
$$
For discrete spectrum an average square momentum is

$$
p_0=\sqrt{-2mE} 
$$
Fock introduces coordinates of a stereographic projection of a unit sphere
in fourdimensional Euclidean space

\begin{eqnarray*}
\xi &=&\frac{2p_0p_x}{p_0^2+{\bf p}^2}=\sin \alpha \sin \vartheta \cos
\varphi \\
\eta &=&\frac{2p_0p_y}{p_0^2+{\bf p}^2}=\sin \alpha \sin \vartheta \sin
\varphi \\
\zeta &=&\frac{2p_0p_z}{p_0^2+{\bf p}^2}=\sin \alpha \cos \vartheta \\
\chi &=&\frac{p_0^2-{\bf p}^2}{p_0^2={\bf p}^2}=\cos \alpha \\
&&
\end{eqnarray*}
so that

$$
\xi ^2+\eta ^2+\zeta ^2+\chi ^2=1 
$$
Schroedinger equation in new coordinates takes the following form

$$
\Psi \left( \alpha ,\vartheta ,\varphi \right) =\frac \lambda {2\pi ^2}\int 
\frac{\Psi \left( \alpha ^{\prime },\vartheta ^{\prime },\varphi ^{\prime
}\right) }{4\sin ^2\frac \omega 2}d\Omega ^{\prime } 
$$
where

$$\lambda =\frac{Zme^2}{h\sqrt{-2mE}}$$

and $2\sin \frac \omega 2$ is distance between points $\alpha ,\vartheta
,\varphi $ and $\alpha ^{\prime },\vartheta ^{\prime },\varphi ^{\prime }$
on four-dimensional unit sphere

$$
4\sin ^2\frac \omega 2=\left( \xi -\xi ^{\prime }\right) ^2+\left( \eta
-\eta ^{\prime }\right) ^2+\left( \zeta -\zeta ^{\prime }\right) ^2+\left(
\chi -\chi ^{\prime }\right) ^2 
$$
The equation for $\Psi \left( \alpha ,\vartheta ,\varphi \right) $ is
nothing but an integral equation for four-dimensional spherical functions.
Function $\Psi \left( \alpha ,\vartheta ,\varphi \right) $ is related to the
wave function in momentum space $\psi \left( {\bf p}\right) $ as follows

$$
\Psi \left( \alpha ,\vartheta ,\varphi \right) =\frac \pi {\sqrt{8}%
}p_0^{-3/2}\left( p_0^2+{\bf p}^2\right) ^2\psi \left( {\bf p}\right) 
$$

Integral equation for $\Psi \left( \alpha ,\vartheta ,\varphi \right) $
contains important information:

1. An invariance group of the Hydrogen-like atom is a group of four
dimensional rotations. This explains why energy levels are independent of
asimutal quantum number and introduces hyperspherical functions in
calculations. The invariance group discovered by Fock is especially useful
in averaging or summation within the layer with given principal quantum
number.

2. In the case of continuous spectrum of Hydrogen-like atoms analogous
approach leads to the Lorentz group symmetry on four-dimensional
hyperparaboloid.

3. The invariance group of the Hydrogen-like atom discovered by Fock is not
a kinematical group which transforms only coordinates or momenta. $\alpha $%
-rotations connect functions with different energy levels $E$ . Such
symmetry was named by Fock  as dynamical. Search for new dynamical groups
became quite popular after 1960, i.e. 25 years after this paper by Fock.

4. Fock has also shown that momentum space of Hydrogen-like atoms is
non-Euclidean. It is endowned by the Riemannien geometry of constant
positive curvature in the case of discrete spectra, and by Lobachevskian
geometry with constant negative curvature in the case of continuous spectra.

\vskip 6pt
{\bf Energy-time uncertainty relation}
\vskip 3pt

Uncertainty relation $\Delta E\Delta t\geq h$ was originally considered as a
relativization of Heisenberg's relations $\Delta x\Delta p_x\geq h$ .
Analysis of Fock and Krylov [12] has shown that energy-time uncertainty
relation should be written as

$$
\Delta (E^{\prime }-E^{})\Delta t\geq h
$$
where  $\Delta \left( E^{\prime }-E\right) $ is uncertainty in energy change
in transition from one state to another, and $\Delta t$ is time for which
transition probability is close to unity. This relation is applicable to any
chosen experiment and results from ''non-absolute description'' of
microsysteme by the wave function.

For quasistationary state $E^{\prime }$ in transition to exactly stationary
state $E$ uncertainty relation defines level width $\Delta E^{\prime }$. If
this width is small so that $\Delta t$ $\cong \tau $ is great, decay
probability is constant in time, and exponential decay follows with lifetime 
$\tau .$ One should remind that behaviour of state during experiment
(i.e.measurement) cannot be described by Schroedinger equation. Uncertainty
in energy change in a system means that uncertainty of interaction energy of
a system in transition grows when time of change shortens.

\vskip 6pt
{\bf A possible generalization of the concept of physical space}
\vskip 3pt

Fock writes [13]:
"The concept of physical space is closely related with that of 
the motion of a physical body. This connection is quite natural since we learn 
the properties of space by studying the motion of physical bodies. .. And the physical
space (as distinguished from the configuration space) was always thought of as a
manifold connected with the degrees of freedom of asingle mass point. 
In quantum physics the simplest kind of physical object (a material point) may be 
supposed to have the properties of an electron."
Fock describes spin degree of freedom of electron and the Pauli principle  and 
concludes:
"Applying to this case the (tacitly assumed) classical presumption that the physical 
space is defined by totality of variables describing the degrees of freedom of the 
simplest physical body (conventionally called a mass point), ...we arrive at the 
generalization of the concept of physical space. The generalized physical space 
defined above can be called spinor space."  The Pauli principle corresponds to the
impossibility for two electrons to occupy one and the same point in spinor space.

This Fock paper may be considered as a physical motivation for
Supersymmetry [14].

\vskip 6pt
{\bf References}
\vskip 3pt

1. Fock V.A.-Zs.f.Phys., 1926, vol.38, 242. \\
2. Fock V.A. - Zs.f.Phys., 1926, vol.39, 226. \\
3 .Klein O. - Zs.f.Phys., 1926, vol. 37, 895. \\
4. Gordon W.- Zs.f.Phys., 1926,vol.40, 117. \\
5. FockV.A. - Zs.f.Phys., 1926, vol. 57, 261. \\
6. FockV.A.- Phys.Zs.Sowet., 1937, vol.3, 404. \\
7.Weisskopf V.- Zs.f.Phys.,1934, vol.89, 27; vol. 90, 817.\\
8. Schwinger J. - Phys. Rev., 1951, vol.82, 664.\\
9. De Witt B.S. - in: Relativity, Groups and Topology, eds. De Witt B.S.,
New York, Gordon and Breach, 1965, p.19.\\
10. Itabashi K.- Prog. Theor.Phys., 1981, vol.85, 1423.\\
11. Fock V.A. -Zs.f.Phys.,1935, vol. 98, 145.\\
12. Krylov N.S., Fock V.A.- JETP, 1947,vol.17, 93. \\
13. Fock V.A. Physica Norvegica, 1971, vol.5, 149. \\
14. Golfand Yu.A., Likhtman E.P., JETP Letters, 1971, vol.13, 452.
    Volkov D.V., Akulov V.P., JETP Letters, 1972, vol.16, 621\\

\end{document}